\documentclass[onecolumn, draftclsnofoot, 12pt]{IEEEtran}
\usepackage{amsmath}
\usepackage{cite}
\usepackage{amssymb}
\usepackage{amsfonts}
\usepackage{algorithm}
\usepackage{dsfont}
\usepackage{graphicx}
\usepackage{epsfig}
\usepackage{subfigure}
\usepackage{psfrag}
\usepackage{xcolor}
\usepackage{url}
\usepackage[colorlinks,linkcolor=black,urlcolor=black,anchorcolor=black,citecolor=black,hyperfootnotes=true]{hyperref}
\newtheorem{example}{Example}

\title{Networked Sensing in 6G Cellular Networks: Opportunities and Challenges}
\author{Liang Liu, Shuowen Zhang, Rui Du, Tony Xiao Han, and Shuguang Cui
\thanks{Liang Liu and Shuowen Zhang are with the Department of Electronic and Information Engineering, The Hong Kong Polytechnic University (e-mail: \{liang-eie.liu,shuowen.zhang\}@polyu.edu.hk).}
\thanks{Rui Du and Tony Xiao Han are with Huawei Technologies Co., Ltd (e-mail: \{ray.du,tony.hanxiao\}@huawei.com).}
\thanks{Shuguang Cui is with the School of Science and Engineering (SSE) and Future Network of Intelligence Institute (FNii), The Chinese University of Hong Kong (Shenzhen) (email: shuguangcui@cuhk.edu.cn).}}
\begin{document}

\maketitle \thispagestyle{empty} \vspace{-0.3in}

\begin{abstract}
Radar and wireless communication are widely acknowledged as the two most successful applications of the radio technology over the past decades. Recently, there is a trend in both academia and industry to achieve integrated sensing and communication (ISAC) in one system via utilizing a common radio spectrum and the same hardware platform. This article will discuss the possibility of exploiting the future sixth-generation (6G) cellular network to realize ISAC. Our vision is that the cellular base stations (BSs) deployed all over the world can be transformed into a powerful sensor to provide high-resolution sensing services. Specifically, motivated by the joint encoding/decoding gain in multi-cell coordinated communication, we advocate the adoption of the \emph{networked sensing} technique in 6G networks to achieve the above goal, where the BSs can share their obtained range/angle/Doppler information with each other for jointly estimating the locations and velocities of targets. Key opportunities and challenges for realizing networked sensing in the 6G era will be revealed in this article. Moreover, avenues for future research along this promising direction will also be outlined.
\end{abstract}

\section{Introduction}
It is envisioned that the sixth-generation (6G) cellular network will be commercialized in around 2030. The two key enablers for 6G broadband communication are the ultra-high frequency band, e.g., the millimeter-wave (mmWave) and Terahertz (THz) bands, and the ultra-massive multiple-input multiple-output (MIMO) technique. Interestingly, the above trend of wider bandwidth and larger antenna array makes the communication signals tend to be of ultra-high resolution in the time domain and angle domain. As a result, the 6G cellular network can potentially make use of the communication signals reflected by the targets to estimate their locations and velocities accurately, similar to the radar system. It is thus envisioned that the future 6G cellular network will not only support high-speed broadband communication, but also provide high-resolution sensing services. Such an integrated sensing and communication (ISAC) system is appealing to various applications such as intelligent transportation systems with autonomous cars and smart factories with moving robots, where the environment information needs to be sensed and conveyed to these machines for their safe operation \cite{ISAC1,ISAC2,ISAC3,ISAC4,ISAC5}.

\begin{figure}[t]
\centering
\includegraphics[width=4in]{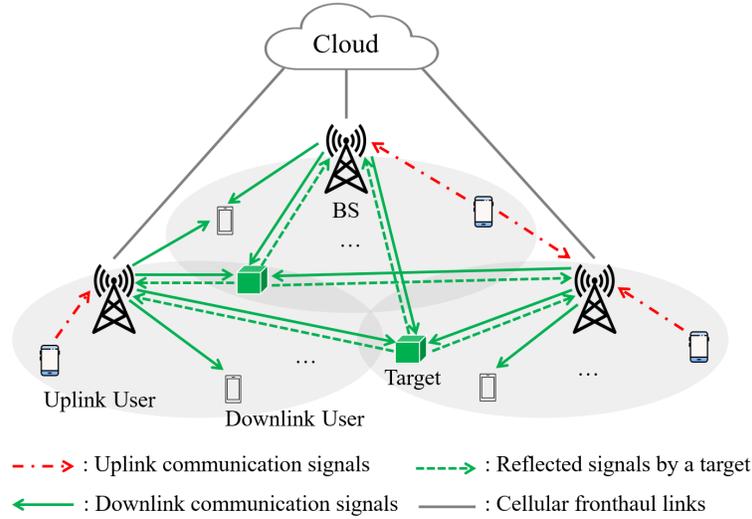}\vspace{-10pt}
\caption{An illustration of the 6G ISAC networks with coordinated communication and networked sensing, where the BSs receive uplink signals from uplink users, and transmit downlink signals to convey information to downlink users and sense the environment simultaneously. Coordinated communication is mature in cellular networks, e.g., an uplink/downlink user can be jointly served by multiple BSs, as shown in the figure. Similarly, under networked sensing, each target may also reflect the downlink signals to multiple BSs, which can share the information of the target extracted from its echoes and jointly estimate the target's location and velocity.}
\label{fig1}\vspace{-10pt}
\end{figure}

Intuitively, it seems that the sensing capability provided by the cellular network can hardly be as powerful as that in the radar system, since the radar system is designed for sensing, while the cellular network is designed for communication. However, there exists a unique property of the cellular network that potentially makes it as competitive as the radar system in sensing. To achieve high-speed communication, the tremendous base stations (BSs) deployed all over the world can collaborate with each other to mitigate the inter-cell interference and even jointly encode/decode the user messages, thanks to the universal cellular standards adopted. Similarly, to achieve high-resolution sensing, they can also collaborate with each other to perform the so-called \emph{networked sensing}, as illustrated in Fig. \ref{fig1}. Specifically, a target may reflect the downlink communication signals to multiple BSs, which can then jointly localize it by sharing the range/angle/Doppler information about this target extracted from its echoes. Note that in practice, each radar system usually works alone for sensing due to the sparsely deployed radar transmitters/receivers and the absence of a unified radar protocol. This motivates us to deeply investigate the potential gain brought by the newly emerging networked sensing technique in this article, which is unique in the cellular network.

\section{Device-Free Sensing in 6G Cellular Networks}

The goal of sensing is to exploit the radio frequency (RF) signals for estimating the locations and velocities of the targets in the environment. In general, sensing can be classified into two categories: device-based sensing and device-free sensing. Specifically, for device-based sensing, the targets are capable of actively transmitting or receiving the RF signals such that their locations and velocities can be estimated based on the one-way signals between the anchors and the targets. For example, the satellites in the Global Navigation Satellite System (GNSS) and the BSs in the current cellular networks can utilize the device-based sensing technique to localize the mobile phones \cite{Device_based_sensing_2017}. On the other hand, for device-free sensing, the targets do not necessarily have to possess the communication capability \cite{Liu_ISAC_2022}. Instead, they merely reflect the signals emitted by the transmitters, while the receivers need to extract the location/velocity information based on the reflected signals. One key application of the device-free sensing technique is the radar system.

Although the device-based sensing technique has been successfully applied in the current cellular networks to localize active devices such as mobile phones, many targets in practice are passive devices without the communication capability, which can only be sensed by the device-free technique. Motivated by the above, in this article, we focus on cellular network based device-free sensing to realize ISAC in the 6G era, which is an open topic despite the tremendous success of the device-free sensing technique in radar systems. To start with, we consider the following two questions that are of paramount importance.

\begin{itemize}
\item[{\bf Q1.}] {\bf Is it \emph{feasible} to utilize the communication signals in cellular networks to achieve device-free sensing as in radar systems?}
\item[{\bf Q2.}] {\bf Is it \emph{trivial} to utilize the communication signals in cellular networks to achieve device-free sensing?}
\end{itemize}

In the following two subsections, we discuss the above two crucial questions, respectively.

\subsection{Opportunities We Can Exploit}
In the fifth-generation (5G) cellular network, mmWave and massive MIMO techniques have built the foundation for high-speed communication, while the trend of increasing the bandwidth and number of antennas will be continued in the 6G network for further improving the communication quality. Fortunately, this will also be beneficial for improving the range and angle resolutions in sensing. For example, the bandwidth of the current 5G mmWave network can be up to $400$ MHz, and the corresponding range resolution is $0.1875$ meters \cite{Liu_ISAC_2022}, i.e., as long as two targets are $0.1875$ meters away from each other, their reflected signals can be separately detected by a BS. Moreover, since the 5G BSs are equipped with tens or even hundreds of antennas, mature angle-of-arrival (AoA) or angle-of-departure (AoD) estimation techniques such as MUSIC can be applied to accurately estimate the targets' angle information. In the future 6G network, it is envisioned that even higher frequency bands, e.g., the THz band, and larger number of antennas, e.g., ultra-massive MIMO, will be adopted to harvest more bandwidth and spatial multiplexing gain. Therefore, we conclude that sensing with ultra-high range and angle resolutions is possible in the future 6G network.

\subsection{Challenges We Should Tackle}\label{sec:challenges}
However, integrating the sensing function in a communication system is not straightforward. Particularly, we view waveform design and duplex technique to cope with the interference from communication signals as the core challenges to realize sensing in 6G networks.

\subsubsection{Waveform Design}
\begin{figure}
\begin{center}
\subfigure[Ambiguity function of Zadoff-Chu based pilot signal]
{\scalebox{0.55}{\includegraphics*{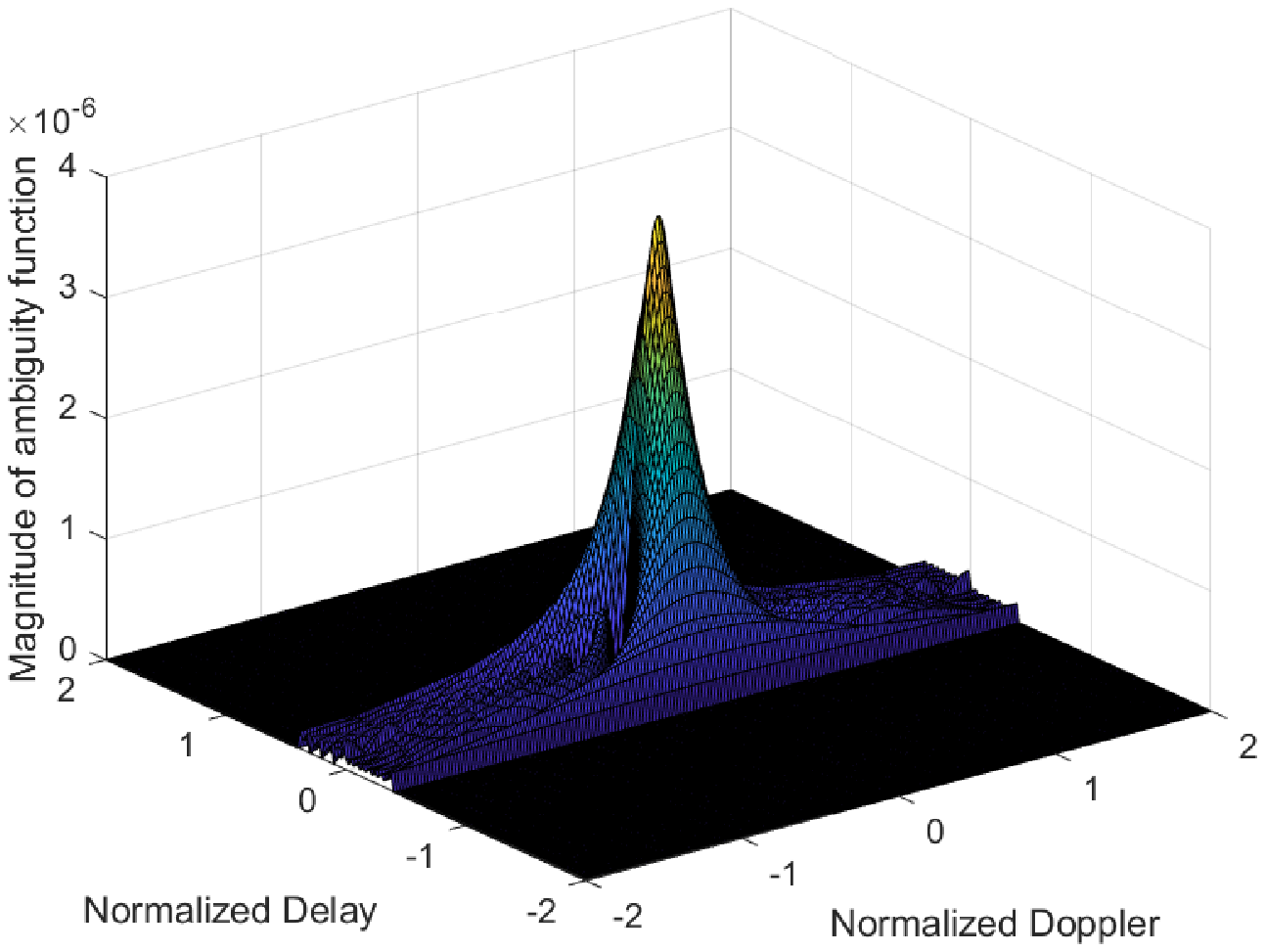}}}
\subfigure[Ambiguity function of OFDM data signal]
{\scalebox{0.55}{\includegraphics*{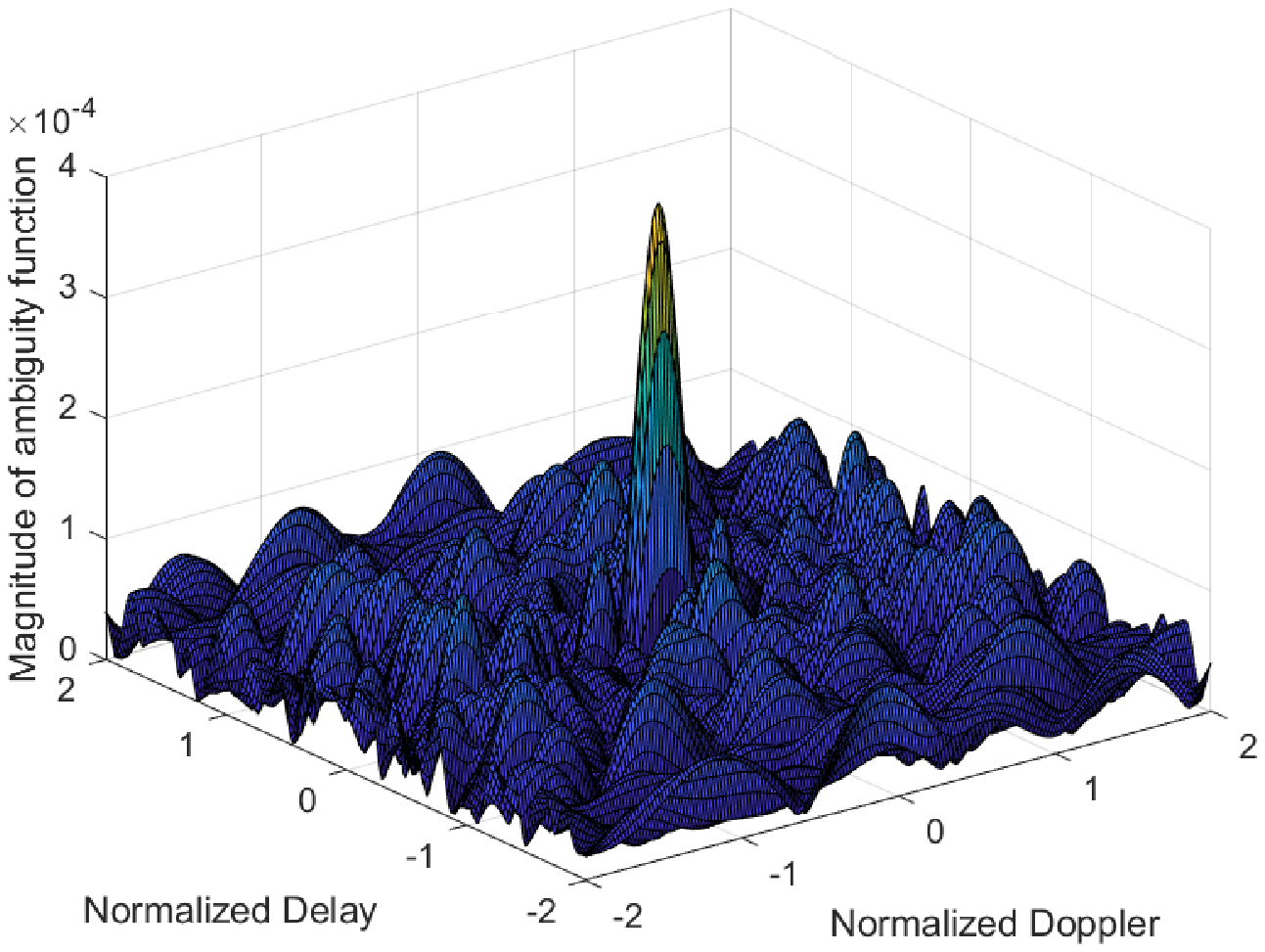}}}
\end{center}\vspace{-10pt}
\caption{Comparison between the ambiguity functions of pilot and data signals.}\label{fig2} \vspace{-15pt}
\end{figure}

In radar systems, the ambiguity function of the designed signal just has a strong main lobe such that the location and velocity of a target can be accurately estimated via a matched filter at the receiver. It is thus interesting to investigate whether the ambiguity function of the communication signals in cellular networks still possess this property. On one hand, the Zadoff-Chu sequence is widely used as the pilot signal in cellular networks for synchronization and channel estimation due to its good correlation property; on the other hand, the orthogonal frequency division multiplexing (OFDM) modulated sequence is widely used as the data signal due to its high spectrum efficiency and ability to combat inter-symbol interference. Fig. \ref{fig2} (a) and Fig. \ref{fig2} (b) show the ambiguity functions of the pilot signal and the data signal in cellular networks, respectively. It is observed that the pilot signal does have very good correlation property, thus can be directly used for sensing. However, there are several strong side lobes in the ambiguity function of the data signal, which is not surprising because the OFDM signal is not dedicatedly designed for sensing. To enable the use of communication data signals for sensing, the waveform of the data signals needs to be carefully designed to mitigate the side lobes in their ambiguity functions. Alternatively, other localization techniques without strict requirements on the ambiguity function of the transmitted signals can also be developed, e.g., the OFDM-based range estimation method adopted in \cite{Liu_ISAC_2022}.

\subsubsection{Duplex Technique}
\begin{figure}[t]
\centering
\includegraphics[width=5in]{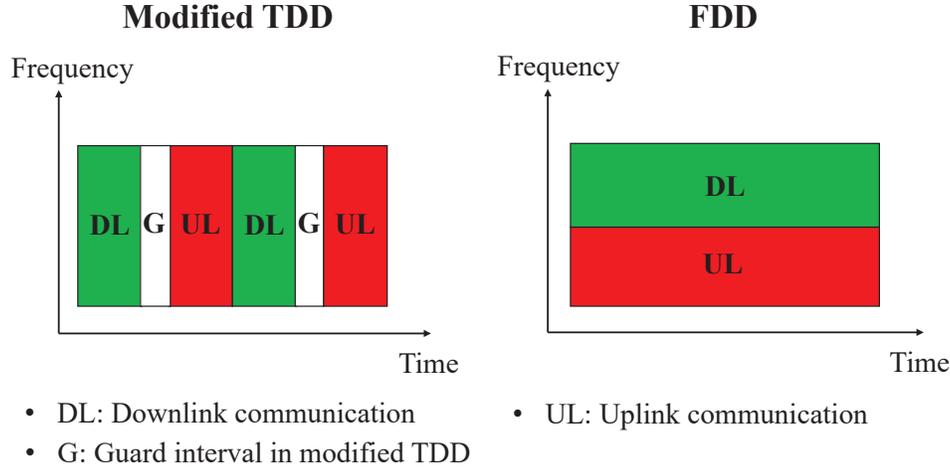}\vspace{-10pt}
\caption{An illustration of the TDD protocol and FDD protocol for ISAC in 6G cellular networks. Under the modified TDD protocol, the downlink signals reflected by the targets are all received before the uplink communication interval starts, thanks to the guard interval added. Under the FDD protocol, the downlink signals reflected by the targets and the uplink signals can be separated in the frequency domain.}
\label{fig3}\vspace{-10pt}
\end{figure}

Besides waveform design, another challenge for sensing in cellular networks is how to mitigate the interference from the communication signals that are not reflected by the targets. Firstly, the reflected signals from the targets to a BS may be interfered by the downlink signals transmitted by the BS itself. As a result, the advanced self-interference cancellation techniques proposed for the full-duplexing systems \cite{Full_Duplexing_2013} should be adopted in ISAC systems. Secondly, the reflected signals from the targets may be interfered by the uplink signals from the mobile users as well. In the current cellular networks, the uplink communication and the downlink communication are separated either in the time domain via the time-division duplexing (TDD) technique or in the frequency domain via the frequency-division duplexing (FDD) technique. We can generalize the above duplex techniques to ISAC systems as follows. For the TDD system, although the downlink signal is transmitted at a different time interval compared to the uplink signal, the downlink signal reflected by some target may arrive at the BS in the same time interval as the uplink signals from the mobile users, thus causing interference. To avoid this issue, the TDD protocol needs to be modified. Specifically, we can add a \emph{guard interval} when a downlink communication interval is transited to an uplink communication interval, as shown in Fig. \ref{fig3}. The duration of the guard interval should be determined by the sensing range of the system such that even the signal reflected by the farthest target can be received by the BS within this interval. For example, if the maximum BS-target distance is $150$ meters, the duration of the guard interval can be set as $10^{-3}$ milliseconds, which is very short as compared to the duration of each uplink/downlink communication interval. Under the modified TDD protocol, the uplink and downlink communication will not interfere with each other, and the sensing performance will not be affected by the uplink communication. On the other hand, for the FDD system, the downlink signals reflected by the targets and the uplink signals from the mobile users are at different frequency bands and will not generate interference to each other. Hence, the FDD protocol can be used in the future ISAC systems without modification, as shown in Fig. \ref{fig3}.

To summarize, it is both theoretically and practically feasible to perform sensing using the communication signals in cellular networks. However, waveform design and duplex technique are of paramount significance to the sensing performance of the future 6G networks. If we can properly handle these issues, one crazy but appealing question arises:
\begin{itemize}
\item[{\bf Q3.}] {\bf Can 6G cellular network achieve the same or even better sensing performance over the existing radar systems?}
\end{itemize}

To answer this question, this article will discuss the new advantages brought by \emph{networked sensing}, which is a unique function provided by the cellular network.

\section{One Step Beyond: Networked Device-Free Sensing in 6G Systems}
The key idea of networked sensing is to transform a large number of BSs in the cellular network into a huge sensor for improving the sensing capability. Note that there is a fundamental difference between radar systems and communication systems: the radars usually work independently due to their sparse deployment and the absence of a globally unified standard; while the BSs can cooperate with each other to manage inter-cell interference and even jointly encode/decode user messages due to their dense deployment and the same standard adopted. Thanks to the above advantage of the cellular network, we can exploit the large-scale cooperation among the BSs to enable networked sensing, where the adjacent BSs share their estimated range, angle, and Doppler information in order to jointly estimate the locations and velocities of the targets, as shown in Fig. \ref{fig1}. Such BS cooperation gain in networked sensing can potentially lead to high-resolution sensing in the future 6G cellular network. It is worth noting that from the data sharing perspective, it is practically easier to implement cooperative sensing than cooperative communication. Specifically, cooperative communication requires the BSs to share a significant amount of information about the channel state information (CSI) and/or the user data with low latency (before the CSI and/or the data become outdated), for which several data compression techniques have been proposed to facilitate the efficient data flow over the fronthaul/backhaul networks that connect the BSs and the cloud processor \cite{CRAN_2016}. Fortunately, for networked sensing, only a very small amount of data traffic about the range, angle, and Doppler information of the targets needs to be shared with a moderate delay requirement. Hence, the current fronthaul/bachkaul techniques to support cooperative communication are sufficient for networked sensing.

Note that in the cellular network, networked device-based sensing has been widely employed to localize active targets such as the mobile phones \cite{Device_based_sensing_2017}. Along this line, typical techniques include the range-based trilateration method, where the distances of a target to three BSs should be known such that its location can be estimated as the unique intersection point of three circles; and the angle-based triangulation method, where the AoAs or the AoDs of a target from/to two BSs should be known such that its location can be estimated as the unique intersection point of two lines; etc. Moreover, the sensing signals among the targets have also been utilized to improve the performance of networked device-based sensing in \cite{Networked_Sensing_2009}. Under the above networked device-based sensing techniques, the range/angle information is extracted from the \emph{one-way} signals transmitted between the \emph{active} targets and the BSs/targets. On the other hand, it is worth noting that under the networked device-free sensing techniques, the range/angle information should be obtained from the \emph{round-trip} signals reflected by the \emph{passive} targets to the BSs. In the rest of this section, we will focus on two unique challenges for networked device-free sensing stemming from the above difference: the coverage issue and the data association issue.

\subsection{BS Sensing Coverage}\label{sec:BS Sensing Coverage}
To enable networked sensing in 6G cellular networks, each target should be detected by multiple BSs. \emph{Coverage}, however, may become an issue for device-free sensing, because localization is performed based on the round-trip reflected signals, which are much weaker than the one-way signals used in device-based sensing. Consequently, to examine the feasibility of networked device-free sensing, we need to carefully think about the following question.

\begin{itemize}
\item[{\bf Q4.}] {\bf What is the maximum coverage range of a BS such that the strength of the reflected signals from the targets in this area is always above a minimum threshold for accurate sensing?}
\end{itemize}

It is known that for wireless communication, the coverage range of a BS can be beyond $1$ kilometer such that the cell-edge users have satisfactory signal-to-noise ratio (SNR) for information decoding. In the following, we discuss the coverage range of a BS required by sensing, which can be very different from that required by communication. Specifically, according to the radar range equation \cite{Radar_1980}, the SNR for device-free sensing can be expressed as
\begin{align}\label{eqn:SNR radar}
\gamma_{{\rm sensing}}=\frac{P_tG_tG_rG_p\lambda^2\sigma_{{\rm rcs}}}{(4\pi)^3kT_0BN_fR^4},
\end{align}where the meaning and some typical value of each notation are listed in Table \ref{table1}. Suppose that the minimum SNR requirement for sensing is $\gamma_{{\rm sensing}}=10$ dB. Then, it can be shown that the BS coverage ranges for sensing a pedestrian and a vehicle are $R=413$ meters and $R=1744$ meters, respectively, which are satisfactory.

\begin{table}
\centering
\begin{center}
\caption{Meanings and Values of Notations in (\ref{eqn:SNR radar})} \label{table1}
{\small
\begin{tabular}{|c|c|c|}
\hline Notation & Meaning & Value \\
\hline
$P_t$ & Transmit Power & $10$ Watts \\
\hline
$G_t$ & Transmit Antenna Gain & $20$ dBi \\
\hline
$G_r$ & Receive Antenna Gain & $20$ dBi \\
\hline
$G_p$ & Processing Gain & $10$ dB \\
\hline
$\lambda$ & Signal Wavelength & $0.086$ meters (at 3.5 GHz) \\
\hline
$\sigma_{{\rm rcs}}$ & Radar Cross Section (RCS) & $-10$ ${\rm dBm}^2$ for pedestrian, $15$ ${\rm dBm}^2$ for vehicle \\
\hline
$k$ & Boltzmann Constant & $1.38\times 10^{-23}$ Joule/Kelvin Degree \\
\hline
$T_0$ & Temperature & $290$ Kelvin Degree \\
\hline
$B$ & Bandwidth & $100$ MHz \\
\hline
$N_f$ & Noise Factor & $5$ dB \\
\hline
$R$ & Distance Between BS and Target & To be calculated based on SNR requirement \\
\hline
\end{tabular}
}
\end{center}
\end{table}

For enlarging the coverage range to achieve better performance of networked sensing, we may increase the transmit power of the BSs, make use of the multi-antenna techniques to increase the antenna gain, or increase the integration time to improve the processing gain. On the other hand, a dense deployment of the BSs is also promising to make each target covered by multiple BSs. For example, the cloud radio access network (C-RAN) with a large number of low-cost remote radio heads (RRHs) \cite{CRAN_2016} can be a good candidate for ISAC. Moreover, recently, \cite{IRS_sensing_2022} proposed a novel networked sensing architecture with heterogeneous anchors, where the low-cost intelligent reflecting surfaces (IRSs) \cite{IRS_2019,Zhang_2020} can be densely deployed to serve as passive anchors in cooperative localization. More details about this topic can be found in Section \ref{sec:Future Directions}. Lastly, we may also utilize the mobile phones as anchors (suppose that their locations can be accurately estimated via device-based sensing) to help sense the area that is not covered by too many BSs.

\subsection{Illustration of the Data Association Issue}
Besides coverage, another issue arising from the passiveness of targets is \emph{data association} \cite{Liu_ISAC_2022}. Specifically, in device-based sensing, the range/angle/Doppler information is estimated based on the one-way signals from/to the active targets to/from the anchors. In this case, by letting different transmitters send signals with different signatures, each receiver can easily match each received signal and the corresponding sensing information (i.e., range/angle/Doppler) contained therein with the right transmitter. However, in 6G-enabled networked device-free sensing, each BS needs to extract the range/angle/Doppler information from the round-trip signals that are reflected by the targets. From each BS's perspective, its transmitted signal is reflected by multiple targets back to itself with the same signature, making it non-trivial to match each reflected signal and the corresponding sensing information contained therein with the right target. In this regard, data association, which is a common challenge for multi-source multi-target sensing \cite{Data_Association_2007}, becomes a critical factor that potentially limits the performance of networked device-free sensing. Note that the data association issue exists in matching different types of sensing information extracted from the received signals with the targets. However, due to the space limitation, we focus our discussion in this article on the data association issue in matching the range (distance) information considering the trilateration-based localization method, where the coordinate of each target is estimated based on its matched distances to all the BSs. In the following, we provide an example to demonstrate why data association may be an issue for networked device-free sensing.

\begin{figure}
\begin{center}
\subfigure[Network topology with ghost targets]
{\scalebox{0.58}{\includegraphics*{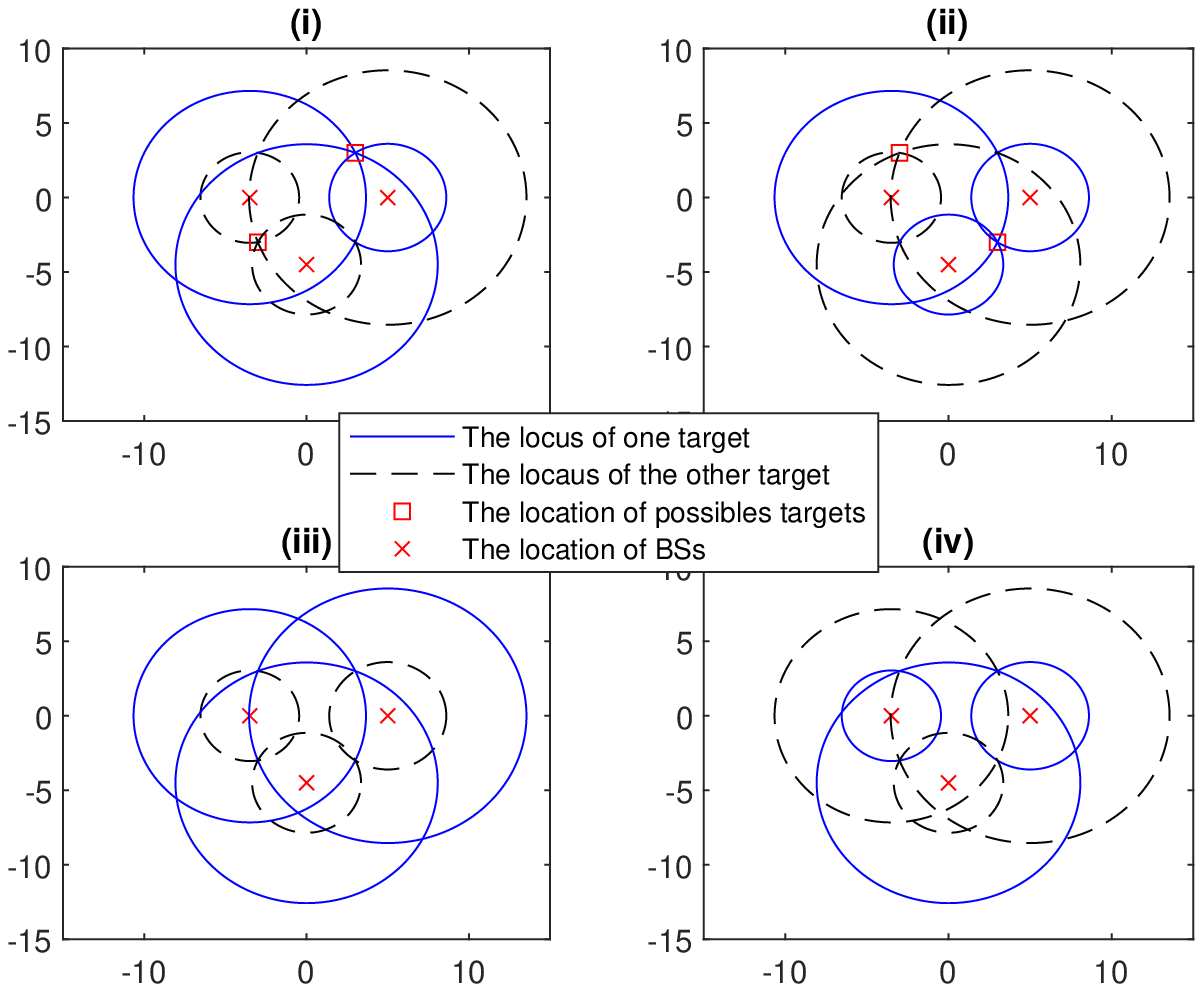}}}\hspace{-25pt}
\subfigure[Network topology without ghost targets]
{\scalebox{0.58}{\includegraphics*{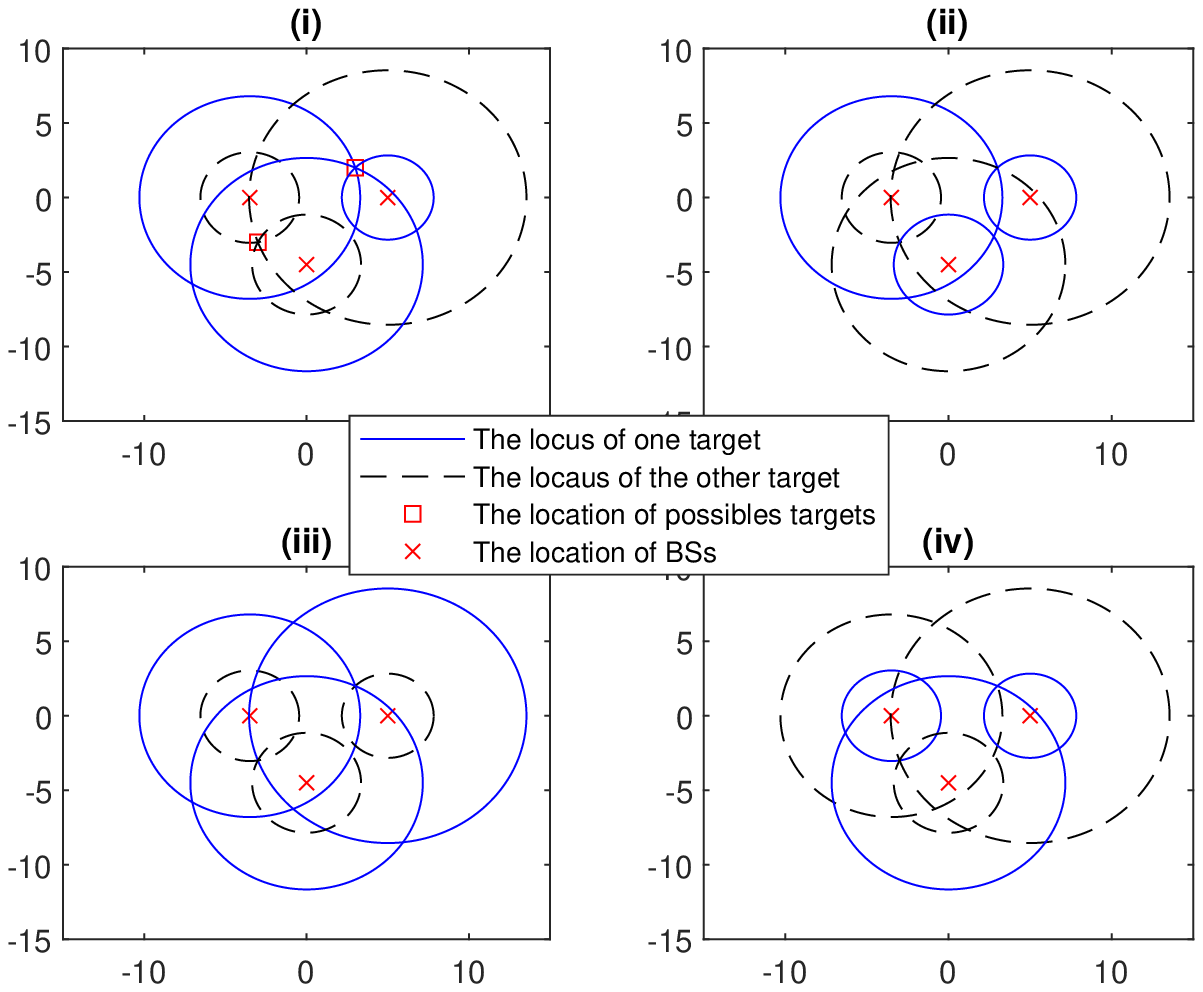}}}
\end{center}\vspace{-10pt}
\caption{Examples with and without ghost targets stemming from wrong data association. Coordinates of the three BSs are $(-3.5,0)$, $(5,0)$, and $(0,-4.5)$. Coordinates of the two targets are $(3,3)$ and $(-3,-3)$ in (a), and $(3,2)$ and $(-3,-3)$ in (b).}\label{fig4} \vspace{-15pt}
\end{figure}

\begin{example}\label{example1}
Consider a networked device-free sensing system as shown in Fig. \ref{fig4} (a), which consists of three BSs with coordinates $(-3.5,0)$, $(5,0)$, and $(0,-4.5)$, and two targets with coordinates $(3,3)$ and $(-3,-3)$. If the range estimation is perfect,  BSs 1, 2, and 3 will respectively collect a distance set of $\{\sqrt{51.25},\sqrt{9.25}\}$, a distance set of $\{\sqrt{13},\sqrt{73}\}$, and a distance set of $\{\sqrt{65.25},\sqrt{11.25}\}$, with the targets.
\end{example}

In Example 1, if BSs 1, 2, and 3 respectively associate the distances $\sqrt{51.25}$, $\sqrt{13}$, and $\sqrt{65.25}$ with localizing target 1, and $\sqrt{9.25}$, $\sqrt{73}$, and $\sqrt{11.25}$ with localizing target 2, the coordinates of these two targets can be correctly estimated as $(3,3)$ and $(-3,-3)$ by applying the trilateration method, as shown in Fig. \ref{fig4} (a-i), indicating that this is a feasible and correct data association solution. However, if BSs 1, 2, and 3 respectively match $\sqrt{51.25}$, $\sqrt{13}$, and $\sqrt{11.25}$ for localizing target 1, and $\sqrt{9.25}$, $\sqrt{73}$, and $\sqrt{65.25}$ for localizing target 2, the coordinates of these two targets will be estimated as $(3,-3)$ and $(-3,3)$, as shown in Fig. \ref{fig4} (a-ii), which are different from the real target coordinates. Thus, this is a feasible but wrong data association solution. Lastly, under other data association solutions, it can be shown that the trilateration method will not lead to any feasible location estimation for the two targets, as shown in Fig. \ref{fig4} (a-iii) and Fig. \ref{fig4} (a-iv), which indicates that they are infeasible data association solutions. To summarize, given the distance set obtained by each BS shown in Example \ref{example1}, there are two feasible data association solutions leading to two feasible location estimations, from which the BSs cannot find the correct target locations. To characterize this effect, we define the false targets at $(3,-3)$ and $(-3,3)$ as the so-called \emph{ghost targets} arising from the wrong data association solution.

The above example indicates that ghost targets may exist in networked device-free sensing due to the existence of multiple feasible data association solutions. However, it is also worth noting that ghost targets do not always exist, as shown in the following example.

\begin{example}\label{example2}
Consider a networked device-free sensing system as shown in Fig. \ref{fig4} (b), where the coordinates of the three BSs and the second target remain the same as in Example \ref{example1}, while the coordinate of the first target is changed to $(3,2)$.
\end{example}

In Example 2, it can be shown that other than the correct data association solution, no other feasible data association solution leading to the detection of two targets exists, as shown in Fig. \ref{fig4} (b). Similar observations were also made in \cite{Liu_ISAC_2022}. Therefore, it is concluded from Examples \ref{example1} and \ref{example2} that the data association issue in networked device-free sensing may or may not result in the detection of ghost targets, depending on the locations of the BSs and the targets.

\subsection{Data Association is Not a Fundamental Limitation, but A Practical Challenge}\label{sec:Why Data Association is Not a Fundamental Limitation}
Motivated by the above, we aim to further explore whether the performance of networked device-free sensing is fundamentally limited by the ghost targets brought by the data association issue, by considering the following question.

\begin{itemize}
\item[{\bf Q5.}] {\bf What is the probability for the ghost targets to exist in networked device-free sensing?}
\end{itemize}

It seems from Example \ref{example1} that wrong data association is inevitable in networked device-free sensing, as long as the locations of BSs and targets satisfy certain conditions. Somehow surprisingly, however, it was recently shown in \cite{Liu_ISAC_2022} that if the range estimation at each BS is perfect for all targets, ghost target arising from wrong data association solution does not exist \emph{almost surely} when the targets are randomly located in the network, as long as any three BSs are not deployed on the same line. This result reveals two facts for randomly located targets. Firstly, given some target location pattern such as the one shown in Example \ref{example1}, ghost targets arising from wrong data association solutions do exist. Secondly, the probability for targets to form a location pattern leading to a wrong data association solution, however, is zero, i.e., almost all the target location patterns result in a unique feasible data association solution, although counter examples do exist with a negligible probability. This is analogous to the philosophy of using the joint typicality property to prove the achievability part of the Shannon capacity: there exist codewords that cause decoding errors; however, the probability of such events vanishes in theory.

The amazing result shown above indicates that theoretically speaking, data association induces no fundamental cost for localization, since from a statistical perspective, we will not face a setup with multiple feasible data association solutions as in Fig. \ref{fig4} (a) with probability one, if the targets are randomly distributed (which is usually the case in practice). This thus lays a solid theoretical foundation for networked device-free sensing in future 6G cellular networks. However, we still face the following issue when implementing networked device-free sensing in practice.

\begin{itemize}
\item[{\bf Q6.}] {\bf How to find the correct data association solution in networked device-free sensing among all the data association solutions?}
\end{itemize}

One straightforward method to find the correct data association solution is to conduct exhaustive search over all solutions. Specifically, given any data association solution, or equivalently, given the associated distances between each target and all BSs, we can estimate the location of each target based on the trilateration method, and record the gaps between the associated target-BS distances and the distances calculated based on the estimated target locations. According to the above theoretical result on the uniqueness of feasible data association solution, there should be only one solution that leads to accurate location estimation of all targets with negligible distance error (gap) with probability one, which is exactly the correct data association solution. However, the computational complexity of the exhaustive search method may be high or even unaffordable in practice. Thus, efficient data association algorithms of lower complexity should be designed to quickly and accurately find the correct data association solution in real-world networked device-free sensing systems. Some attempts in this realm can be found in \cite{Liu_ISAC_2022}.

\section{Future Directions}\label{sec:Future Directions}
Networked device-free sensing is a promising solution to realize ISAC in 6G cellular networks. Along this direction, there are still many open problems worth studying from the perspectives of fundamental theory and practical implementation.
\subsection{Fundamental Theory}
In this article, we have discussed the theoretical foundation of range-based networked device-free sensing \cite{Liu_ISAC_2022}. Thanks to the massive MIMO technique, BSs in the cellular network are usually equipped with a large number of antennas and are able to estimate the angle information of the targets accurately. It is thus interesting to investigate the fundamental limit of angle-based networked device-free sensing. Moreover, the fundamental limit of Doppler estimation is also worth pursuing for the case with mobile targets (e.g., vehicles).

\subsection{Practical Implementation}
\subsubsection{Waveform Design for ISAC}
A critical practical factor that determines the performance of (networked) sensing over cellular networks is the waveform design. As shown in Fig. \ref{fig2}, the pilot signals possess much better correlation properties than the data signals. However, in a practical multi-cell system, it is difficult to force all the BSs to transmit the pilot signals for channel estimation at the same time. In other words, there are always BSs that are transmitting data signals when networked sensing is performed. This thus motivates the study on the waveform design of data signals for better sensing performance.

\subsubsection{Non-Line-of-Sight (NLoS) Path Mitigation}
In practice, another major factor limiting the performance of networked device-free sensing is multi-path propagation, i.e., the existence of NLoS paths from a BS to a target back to the BS via one or more scatters present in the environment. Since the NLoS paths usually do not provide useful information about the target location and velocity, the techniques to mitigate them have been widely studied in device-based sensing systems. However, novel NLoS path mitigation solutions are still needed under the device-free sensing setup.

\subsubsection{Networked Sensing with Heterogeneous Anchors}
\begin{figure}[t]
\centering
\includegraphics[width=3in]{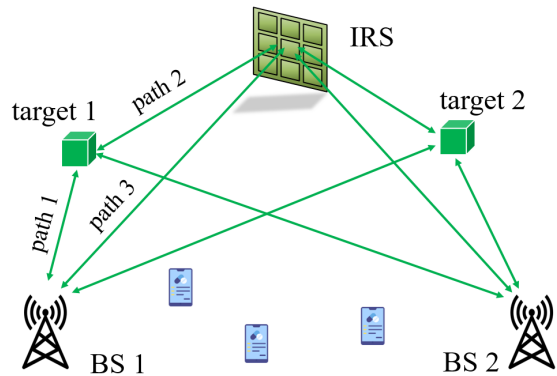}\vspace{-10pt}
\caption{An illustration of the networked device-free sensing architecture with heterogeneous anchors (two BSs and one IRS in this example). Due to the passiveness of the IRS and the targets, the distance between target 1 and the IRS as well as that between target 2 and the IRS cannot be directly estimated by the signals propagated between them. However, \cite{IRS_sensing_2022} showed that these distances can be estimated indirectly in a smart way. For example, for target 1, the length of path 1 and the sum of the lengths of paths 1-3 can be respectively extracted from the signal from BS 1 via target 1 to BS 1 and that from BS 1 via target 1 and the IRS to BS 1. Moreover, the length of path 3 is also known because the locations of BS 1 and the IRS are known. Hence, the distance between target 1 and the IRS can be obtained by subtracting the length of path 1 and the length of path 3 from the sum of the lengths of paths 1-3. After the distances from target 1 to the two active anchors, i.e., BSs 1 and 2, and the passive anchor, i.e., IRS, are known, its location can be estimated based on the trilateration method.}
\label{fig5}\vspace{-10pt}
\end{figure}
As discussed in Section \ref{sec:BS Sensing Coverage}, BS coverage range is an essential concern for networked sensing. Recently, there has been a flurry of research activities in using IRS to enhance the communication range in the 6G era \cite{IRS_2019,Zhang_2020}. From the sensing perspective, it is also appealing to densely deploy the low-cost IRSs to make each target covered by multiple BSs/IRSs. Motivated by this, \cite{IRS_sensing_2022} proposed a novel networked device-free sensing architecture with heterogeneous anchors, i.e., the active BSs and the passive IRSs, as illustrated in Fig. \ref{fig5}. We believe that such a novel architecture is promising for the future ISAC system and should be investigated in more depth.

\section{Conclusions}
In this article, we identified several opportunities and challenges for achieving networked device-free sensing in the future 6G cellular network. Our vision is to transform the BSs in cellular networks into a powerful sensor for sensing the environment with high resolution. It is hoped that this article can motivate more research endeavour in this promising area, for paving the way to the future integration of sensing and communication in a unified system.

\end{document}